\documentclass[apj]{emulateapj}

\usepackage{soul,color,amsmath}

\shorttitle{Migration of Gap-Opening Planets}
\shortauthors{Duffell et al}

\begin{document}

\author{Paul C.~Duffell\altaffilmark{1},
Zoltan~Haiman\altaffilmark{2}, Andrew I.~MacFadyen\altaffilmark{1}, Daniel J.~D'Orazio\altaffilmark{2} and Brian D.~Farris\altaffilmark{1,2}}
\altaffiltext{1}{Center for Cosmology and Particle Physics, New York University}
\altaffiltext{2}{Department of Astronomy, Columbia University}
\email{pcd233@nyu.edu}

\title{The Migration of Gap-Opening Planets is not Locked to Viscous Disk Evolution}

\begin{abstract}

Most standard descriptions of Type II migration state that massive, gap-opening planets must migrate at the viscous drift rate.  This is based on the idea that the disk is separated into an inner and outer region and gas is considered unable to cross the gap.  In fact, gas easily crosses the gap on horseshoe orbits, nullifying this necessary premise which would set the migration rate.  In this work, it is demonstrated using highly accurate numerical calculations that the actual migration rate is dependent on disk and planet parameters, and can be significantly larger or smaller than the viscous drift rate.  In the limiting case of a disk much more massive than the secondary, the migration rate saturates to a constant which is sensitive to disk parameters and is not necessarily of order viscous rate.  In the opposite limit of a low-mass disk, the migration rate decreases linearly with disk mass.  Steady-state solutions in the low disk mass limit show no pile-up outside the secondary's orbit, and no corresponding drainage of the inner disk.

\end{abstract}


\section{Introduction}
\label{sec:intro}

Disk-mediated migration is an important process in the formative years of a planetary system (for a review, see \cite{2012ARAnA..50..211K}).  It can also be important for the dynamical evolution of binary systems including supermassive black hole binaries, which can be driven to merge by a circumbinary disk \citep{2012AdAst2012E...3D}.  The current theoretical understanding of this migration process is incomplete; one point of tension is that planet-disk theory predicts migration timescales much shorter than the lifetime of the protoplanetary disk, a fact which is at odds with the large abundance of planetary systems \citep{2008ApJ...673..487I, 2009AnA...501.1139M, 2009AnA...501.1161M}.  This fast migration is a problem both for low-mass planets, where the interaction is linear (Type I migration, \cite{1980ApJ...241..425G, 1986Icar...67..164W}), but it is also a significant problem for Jupiter-mass planets, which open a gap in the disk (Type II migration, \cite{1986ApJ...309..846L, 1997Icar..126..261W}).

In fact, the Type II regime may be the largest source of tension due to the simplicity and seeming inevitability of the arguments for calculating the Type II migration rate \citep{2013ApJ...774..146H}.  If torques from the planet act as a dam preventing gas from crossing, it becomes inevitable that migration is locked to the viscous evolution of the disk.  If the planet migrated too quickly, it would collide with the inner disk, and leave the outer disk behind.  If it were to migrate too slowly, the outer disk would pile up as the inner disk drained away.

However, this narrative is dependent upon the assumption that gas never crosses the gap.  It turns out that this premise is incorrect; gas can easily move past the planet on horseshoe orbits.  For example, if the planet is held fixed at some orbital position, the inner disk does not drain away; it is completely replenished by gas from the outer disk.  Type II migration may be an idealized situation which requires that there be no gas flowing between the inner and outer disk.  Even if a planet opens a gap, this might not be sufficient for it to migrate at the Type II rate.  Therefore, the migration rate of gap-opening planets must generically depend on the planet mass, the local disk density, the Mach number, and possibly on thermodynamic properties of the disk, such as the timescales for radiative heating and cooling \citep{2006AnA...459L..17P}.

Previous studies have already shown that gas can cross the gap.  This is true for binaries surrounded by a circumbinary disk \citep{1996ApJ...467L..77A, 2008ApJ...672...83M, 2014ApJ...783..134F} as well as the extreme mass ratio case, i.e. planets and extreme mass-ratio black hole binaries \citep{1999ApJ...526.1001L, 1999MNRAS.303..696K, 2001MNRAS.320L..55M, 2003MNRAS.341..213B, 2006ApJ...641..526L, 2007AnA...461.1173C, 2014ApJ...782...88F}.  Conversely, it has been shown in the extreme mass ratio case that gravitational wave torque forcing the secondary in quickly through the inner disk does not have a ``snow plow" effect; the secondary can overtake gas, which moves from the inner disk to the outer disk on horseshoe orbits \citep{2012MNRAS.423L..65B}.

It is also already known that a low disk mass can lead to slower migration \citep{1995MNRAS.277..758S, 1999MNRAS.307...79I}.  This dependence on disk mass has also been confirmed in several numerical studies \citep{2007ApJ...663.1325E, 2008arXiv0807.0625E, 2007MNRAS.377.1324C, 2008ApJ...685..560D}.  However, many current analytical treatments of migration in this case still assume no mass crosses the gap.  In this picture the planet's migration rate increases as mass slowly piles up in the outer disk to drive the planet inward.  Other analytical treatments allow for the possibility of gas crossing the gap \citep{2012MNRAS.427.2680K, 2012MNRAS.427.2660K, 2013ApJ...774..144R}, yet the calculation is still one-dimensional, and therefore any gas which crosses the gap must be imposed in some ad-hoc fashion.  If gas easily crosses the gap, it will imply slower migration rates in low-mass disks because there is no pile-up in the outer disk.  It also allows for the possibility of a gap-opening planet to migrate outward as the disk migrates inward, which is not natural in the standard Type II theory.

Given all of this, one might wonder whether planet migration is tied to the viscous rate in any regime.  Gap-opening planets which allow gas to cross the gap might not be subject to the predictions of Type II migration.  While many previous studies have found that the migration rate is not exactly equal to the viscous rate, most such studies, when using a disk mass comparable to or greater than the mass of the secondary, have found migration rates of order the viscous rate.  However, the space of disk and planet parameters has not been extensively explored, and therefore the fact that this measured migration timescale is of order the viscous timescale may be coincidence.  So far, the most systematic studies of giant planet migration appear to be those of \cite{2007MNRAS.377.1324C} and \cite{2007ApJ...663.1325E, 2008arXiv0807.0625E}, but even these studies did not vary the disk aspect ratio, $h/r$ or equivalently the Mach number $\mathcal{M}$.

In fact, some cases where gap-opening planets did not migrate at the Type II migration rate have already been discovered by \cite{2007ApJ...663.1325E, 2008arXiv0807.0625E}, who found that migration rates are not necessarily proportional to the viscosity.  Since gas consistently crosses the gap in many (possibly all) of the numerical hydro studies to date, one should not necessarily expect these to satisfy the conditions necessary for Type II migration.

The purpose of this work is first to provide a simple, clear demonstration that gas can easily flow past the planet (Section \ref{sec:cross}).  Secondly, a novel means is presented for numerically calculating a self-consistent migration rate for gap-opening planets (Section \ref{sec:numerics}).  The results of such a calculation are presented in Section \ref{sec:results}, where migration rates both significantly larger and smaller than the viscous drift rate are discovered.  These results are discussed in Section \ref{sec:disc}.

\section{Material Crossing the Gap}
\label{sec:cross}

The first step is to demonstrate that massive planets do not act as ``dams" separating the inner disk from the outer disk.  Rather, material can pass from one side of the gap to the other on horseshoe orbits.  The test problem which demonstrates this clearly is to place a Jupiter-mass planet on a fixed circular orbit, and search for a steady-state solution.  This constitutes the limit where the disk mass is much smaller than the planet mass, so if gas had trouble crossing the gap, the inner disk would drain and the outer disk would pile up.  If gas could not cross the gap, then no steady-state solution would be possible.

For simplicity, the initial disk is set to have uniform surface density $\Sigma_0$, uniform pressure $P$, Keplerian orbital velocity ($\Omega \propto r^{-3/2}$), and constant viscosity $\nu$.  The viscous drift velocity $v_r = -\frac32 \nu/r$ ensures a uniform mass flux through the system, $\dot M = 3 \pi \Sigma_0 \nu$.  The boundary conditions are Dirichlet, enforcing this constant mass flux at all times.

To see clearly how mass in the outer disk can sneak past the planet on horseshoe orbits, a passive scalar quantity $X$ is added to the field equations.  Initially, $X = 0$ for $r < a$ and $X = 1$ for $r > a$, where $a$ is the radial position of the planet.  The passive scalar then indicates which fluid elements started in the outer disk, and which started in the inner disk.

Our canonical set of parameters is chosen such that the planet is just massive enough to satisfy the gap-opening criterion, $\Sigma_{gap} \approx 0.1 \Sigma_0$ (this gap criterion has been used in several other studies, e.g. \cite{2006Icar..181..587C}).  The pressure $P = .0025 \Sigma_0 (\Omega_p a)^2$ is chosen to give a Mach number $\mathcal{M} = 20$ in the vicinity of the planet.  The disk viscosity is chosen to be $\nu = 2.5 \times 10^{-5} a^2 \Omega_p$, corresponding to $\alpha = 0.01$ at the planet's position (In later sections, when the planet is moved, viscosity and pressure are not dependent on the planet position; $a$ and $\Omega_p$ in these formulas are evaluated at the planet's final position).  Due to scale invariance, the choice of $\Sigma_0$ is arbitrary.  The system is evolved using the DISCO code \citep{2012ApJ...755....7D, 2013ApJ...769...41D}, assuming a nearly isothermal equation of state (adiabatic index $\gamma = 1.0001$).

\begin{figure}
\epsscale{1.0}
\plotone{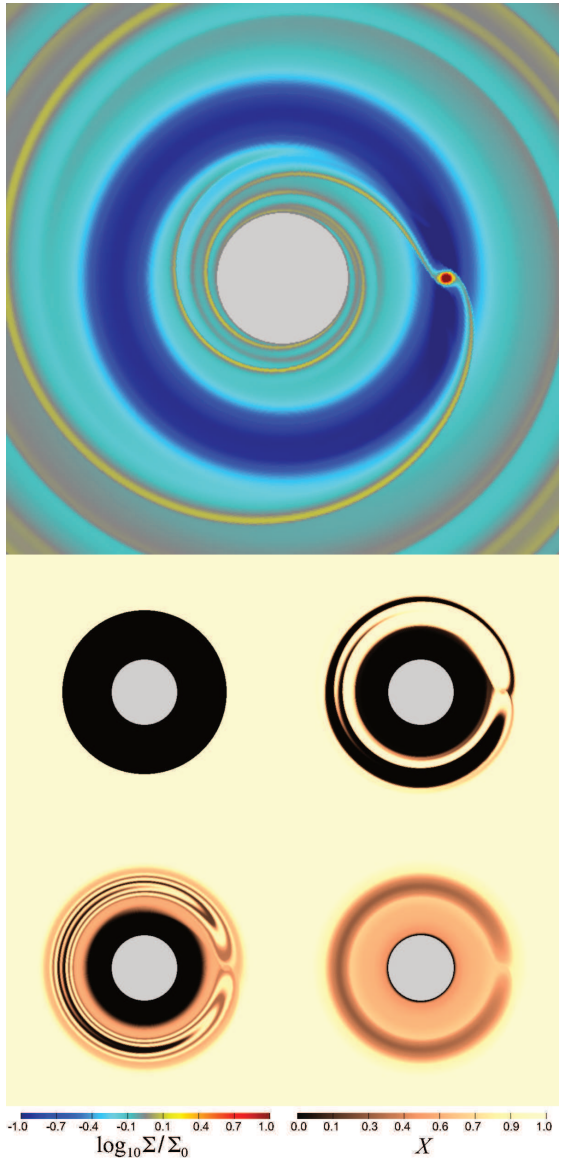}
\caption{ (Top Panel): Logarithm of density for a simple test problem where a Jupiter-mass planet is held on a fixed circular orbit for more than a viscous time (5000 orbits).  The inner disk does not drain away and the outer disk does not pile up.  (Lower Panels): A time series showing gas moving from the outer disk to the inner disk on horseshoe orbits.  A passive scalar is plotted to distinguish material which originated in the outer disk from material that started from the inner disk.  The separate panels show this passive scalar at the initial time, and after 6 orbits, 40 orbits, and 400 orbits.
\label{fig:sneak} }
\end{figure}

The results of this test problem are shown in Figure \ref{fig:sneak}.  The top panel shows the logarithm of density, demonstrating that the planet opens a gap and that a steady-state solution is established after a viscous time ($\sim 4200$ orbits).  The inner disk has not drained away and the outer disk has not piled up.  Actually, the evolution of this disk had already become steady-state after a few hundred orbits, corresponding to Jupiter's gap opening timescale.  In the lower panel, the passive scalar $X$ is plotted, showing the evolution of the inner and outer disk material.  First, gas near $r \approx a$ is pulled into the horseshoe orbits in the corotation region. Soon after the gap forms, gas begins to viscously drift from the outer disk to the corotation region, where it can be pulled past the planet and deposited into the inner disk.  After 400 orbits, the inner disk is nearly completely replaced by material from the outer disk.  After 5000 orbits, the inner disk material is completely gone, and replaced with outer disk material.  This test was also performed with a more massive planet ($3 M_J$), which opens a much deeper gap ($\Sigma / \Sigma_0 \sim 10^{-3}$).  The results for this case were qualitatively the same.

\section{Numerical Strategy}
\label{sec:numerics}

Most numerical studies of the migration of gap-opening planets employ a ``live" planet \citep[e.g.][]{2000MNRAS.318...18N, 2007MNRAS.377.1324C, 2007ApJ...663.1325E, 2008arXiv0807.0625E, 2008ApJ...685..560D}; that is, the planet exerts a gravitational influence on the gas, and simultaneously the planet moves in response to the gravitational influence of the gas.  For a planet with a gap, a self-consistent migrating planet is necessary to capture the details of the previously stated narrative.

In this work, a new strategy is employed, which can complement studies which employ a live planet.  As mentioned above, to study migration of a planet with a gap, it is necessary that the planet be moved, otherwise one will not capture the effect of migration on the disk that the Type II formula is based upon.  However, instead of a ``live" planet, here the planet is moved at a prescribed migration rate, momentarily ignoring the torques exerted on the planet.  In doing this, one eliminates the gravitational influence of the gas, and therefore scale-invariance is restored to the equations.  Thus, the choice of the overall scale of the surface density, $\Sigma_0$, is arbitrary.

After finding a steady-state solution corresponding to a given migration rate (``Steady-state" here is defined to mean that the properly normalized torque plateaus to a constant), it is possible to calculate the time-averaged torques on the planet as it migrates through the disk (torque is averaged over roughly ten orbits, and when properly normalized plateaus to a constant during the numerical integration time of $10^3$ orbits, for all cases studied).  Such torques will be proportional to the choice of $\Sigma_0$, the overall scaling of density.  Therefore, if the sign of the torque is the same as the direction of the migration, there will always exist a self-consistent choice of $\Sigma_0$ which corresponds to the imposed migration rate.

This value of $\Sigma_0$ is calculated for each self-consistent migration rate $\dot a$, which provides a measurement of $\Sigma_0$ as a function of $\dot a$.  This function is then inverted to provide the migration rate as a function of the local surface density, $\dot a ( \Sigma_0 )$.

The technique of prescribing a drift rate to study solutions has been employed before \citep{2003ApJ...588..494M}, but in our case this prescription is used to calculate a self-consistent torque and drift rate, whereas \cite{2003ApJ...588..494M} used this technique to study the stability of solutions.  To describe this more concretely, define the following dimensionless migration rate and normalized torque:

\begin{equation}
w \equiv {\dot a \over \Omega a}
\end{equation}

\begin{equation}
t(w) \equiv \Gamma / \Gamma_0
\end{equation}

where $\Gamma$ is the torque felt by the planet, and it is normalized using $\Gamma_0$, the Type I torque:

\begin{equation}
\Gamma_0 \equiv q^2 \mathcal{M}^2 \Sigma_0 a^4 \Omega^2,
\end{equation}

where $q$ is the planet-to-star mass ratio.  In order for the migration rate to be consistent with the normalized torque calculated,  loss of angular momentum straightforwardly gives the following formula:

\begin{equation}
\dot a = { 2 \Gamma \over m_p a \Omega },
\end{equation}

where $m_p$ is the planet mass.  Combining all of the above formulas gives an expression for the consistent surface density as a function of the chosen migration rate and measured $t(w)$:

\begin{equation}
{\Sigma_0 a^2 \over m_p} = { 1 \over 2 q^2 \mathcal{M}^2 } {w \over t(w)}.
\label{eqn:mig}
\end{equation}

This formula is a function of $w$ which can be inverted to give the migration rate as a function of inner disk mass.

Note that it is possible for more than one migration rate to be consistent with a given $\Sigma_0$, in which case the inverted function is multi-valued.  In other words, this method also provides a means of discovering non-unique migrating solutions, such as Type III migration.

\begin{figure}
\epsscale{1.0}
\plotone{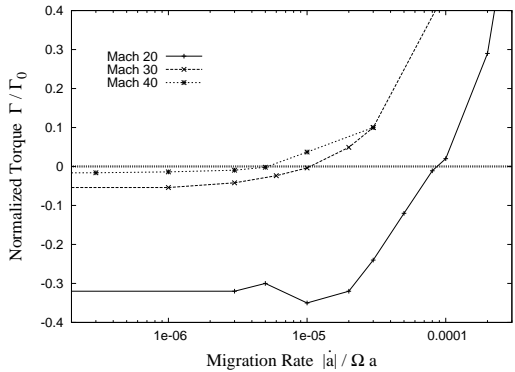}
\caption{ Jupiter is forced to migrate at various rates through the disk, and the torque is measured as a function of the migration rate.
\label{fig:torque} }
\end{figure}

\section{Results}
\label{sec:results}	

\begin{figure}
\epsscale{1.0}
\plotone{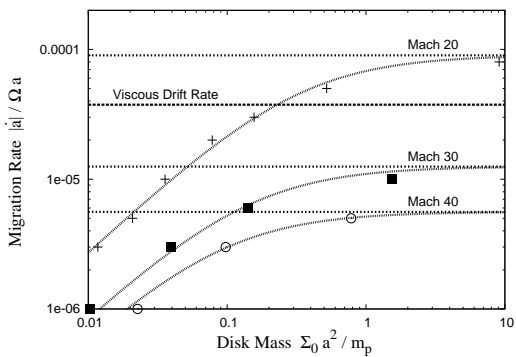}
\caption{ Information from Figure \ref{fig:torque} is used to calculate a self-consistent migration rate as a function of the disk mass, using Equation (\ref{eqn:mig}).  Also plotted is a fitting function given by Equation (\ref{eqn:fit}).
\label{fig:mig} }
\end{figure}

For the present work, the planet mass is fixed to be Jupiter, $q = 10^{-3}$.  Jupiter is pushed through the disk at a variety of migration rates, and the resulting normalized torque on the planet is plotted in Figure \ref{fig:torque}.  For our Mach $20$ disk, the torque changes sign at a migration rate of about $w \sim 10^{-4}$, which is about three times the viscous drift rate ($\dot a_{visc} = -(3/2)\nu/a$).  If $\Sigma_0$ is large enough, the planet can migrate at this rate.  It should be noted that this maximum rate is also not necessarily tied to the viscous rate, and in general can depend on disk parameters, as will be discussed below.  For all planets which migrate slower than this timescale, the torque is negative, and therefore it is possible to calculate the surface density $\Sigma_0$ which gives a torque consistent with the prescribed migration rate.

In Figure \ref{fig:mig}, the migration rate is plotted against this calculated disk mass (expressed as a fraction of the planet mass, $a^2 \Sigma_0 / m_p$).  For this Jupiter-mass planet, a wide range of migration rates is possible.  The Mach $20$ results confirm that the migration rate is proportional to disk mass for low-mass disks, as found previously by \cite{2007ApJ...663.1325E, 2008arXiv0807.0625E} and \cite{2007MNRAS.377.1324C}.   For high disk mass, the migration rate asymptotes to its maximum at about three times the viscous rate.  One may be tempted to conclude that this maximum migration rate is some fixed fraction of the viscous drift rate, but this is not necessarily so.  As an example, additional calculations were performed which varied the Mach number, producing much deeper gaps (by simply choosing a smaller pressure, keeping viscosity fixed) and found variations in this maximum drift rate by more than an order of magnitude (Figure \ref{fig:mig}).  At this stage, it is unclear whether this maximum migration rate is actually related to the viscous rate, but it may be set by the gap opening rate, as the Mach 40 gaps are much deeper and take significantly longer to carve out.  This will be checked in a future work.

In Figure \ref{fig:mig}, a fitting function is also plotted which reasonably fits the data.  This fitting function is given by the following:

\begin{equation}
{ |\dot a| \over \Omega a} = w_1 \left( 1 + {w_1 \over w_0} { m_p \over \Sigma_0 a^2 } \right)^{-1}.
\label{eqn:fit}
\end{equation}

The fitting parameters for this calculation were found to be $w_0 = 2.8 \times 10^{-4}$ and $w_1 = 9 \times 10^{-5}$ for Mach number $\mathcal{M} = 20$.  For $\mathcal{M} = 40$, the migration rates are much slower, $w_0 = 6.4 \times 10^{-5}$, $w_1 = 5.6 \times 10^{-6}$.  $w_0$ is calculated in the slow migration limit,

\begin{equation}
w_0 = 2 q^2 \mathcal{M}^2 | t(0) |,
\end{equation}

and $w_1$ is the migration rate at which the torque changes sign, $t(w_1) = 0$.  It is unclear how exactly these parameters scale with planet mass, Mach number, or viscosity.  In the limited study thus far, the maximum migration rates $w_1$ appear to have a steep scaling with Mach number.  Also, note that although this formula includes the planet mass, the scaling of the parameters $w_0$ and $w_1$ with $m_p$ has not yet been determined, so one should be cautious of over-interpreting the scaling with $m_p$.  This will be determined empirically in a future work.

\section{Summary}
\label{sec:disc}

Type II migration is a fundamental process in astrophysics, and the premise that gas cannot cross the gap has been invoked in many contexts.  This premise is untrue.  A steady mass flux across the gap generically occurs, facilitated by horseshoe orbits in the gap.  This means that gap-opening planets are not destined to migrate at the viscous timescale, but will migrate at a timescale dictated by planet and disk parameters.  This timescale can be significantly longer or shorter than the viscous timescale.

This motivates a more comprehensive study of planet migration in the nonlinear gap-opening regime, as a function of the planet mass and various disk parameters.  This is intended in a future work.  These results should also be checked by live-planet calculations which examine a large range of disk parameters, also examining the effect of different choices of boundary conditions, to see for example whether the choice of Dirichlet vs. outflow boundary conditions has a significant impact on the solution.

\acknowledgments
This research was supported in part by NASA through grant NNX11AE05G issued through the Astrophysics Theory Program.

Resources supporting this work were provided by the NASA High-End Computing (HEC) Program through the NASA Advanced Supercomputing (NAS) Division at Ames Research Center.  We are grateful to Eugene Chiang, Mike Kesden, Andrei Gruzinov, Roman Rafikov, and Scott Tremaine for helpful comments and discussions.  We would also like to thank the anonymous referee for his or her thorough and rigorous review.

\bibliographystyle{apj}

\end{document}